\documentclass[sigconf]{acmart}

\AtBeginDocument{%
  \providecommand\BibTeX{{%
    \normalfont B\kern-0.5em{\scshape i\kern-0.25em b}\kern-0.8em\TeX}}}
\usepackage{dblfloatfix}
\usepackage{listings}
\usepackage{fancyvrb}
\usepackage{hyperref}
\usepackage{url}
\usepackage{balance}
\setcopyright{rightsretained}
\begin{document}

\title{\textit{StackEmo}- Towards Enhancing User Experience by Augmenting Stack Overflow with Emojis}

\author{Akhila Sri Manasa Venigalla and Sridhar Chimalakonda}

\affiliation{%
 \institution{Research in Software and Human Analytics (RISHA) Lab}
 \institution{Indian Institute of Technology}
  \city{Tirupati}
 \country{India}    
 }
\email{{cs19d504, ch}@iittp.ac.in}
\begin{abstract}
With the increase in acceptance of open source platforms for knowledge sharing, Question and Answer (Q\&A) websites such as \textit{Stack Overflow} have become increasingly popular in the programming domain. Many novice programmers visit Stack Overflow for reasons that include posing questions, finding answers for issues they come across in the process of programming. Practitioners voluntarily answer questions on Stack Overflow based on their experience or prior knowledge. Most of these answers are also accompanied by comments from users of Stack Overflow. Questions, answers and comments on Stack Overflow also include sentiments of users, which when analysed and presented could motivate users in reading and contributing to the posts. However, the sentiment of these posts is not being depicted in the current Stack Overflow platform. There is extensive research on analysing sentiments on social networking platforms such as twitter. Representing sentiment of a post might motivate users to follow or answer certain posts. While there exist several tools that augment or annotate Stack Overflow platform for developers, we are not aware of tools that deal with sentiment of the posts. In this paper, we propose \textit{StackEmo} as a Google Chrome plugin to augment comments on Stack Overflow with emojis, based on the sentiment of the comments posted, with the aim to provide users with visual cues that could motivate the users to review and contribute to available comments. We evaluated \textit{StackEmo} through an in-user likert scale based survey with 30 university students. The results of the survey provided us insights on improving \textit{StackEmo}, with 83\% participants having recommended the plugin to their peers.   
\end{abstract}

\keywords{Stack Overflow, Emotion Analysis, LDA, Rule Based Classifier model, Emojis}

\maketitle

\section{Introduction}
Open Source platforms have become increasingly popular among various domains such as medicine \cite{schindelin2012fiji}, bio-informatics \cite{le2009fpocket}, programming \footnote{\url{https://www.eclipse.org/}}, software development \footnote{\url{https://developer.android.com/}} and so on since the last few decades. Software such as Eclipse, NetBeans and Android Studio are available as open source to assist programmers and developers in programming and project development \cite{afzal2018study}. In the domain of software engineering, Open Source Code sharing platforms are being widely used by developers and novice programmers \cite{peng2014collaborative, zagalsky2015emergence}. Question and Answer(Q\&A) websites such as Stack Overflow support knowledge sharing by facilitating users to ask and answer questions in the field of programming \cite{barua2014developers}. There are about 18M\footnote{\url{https://data.stackexchange.com/stackoverflow/query/1136062}} questions and 28M\footnote{\url{https://data.stackexchange.com/stackoverflow/query/1136061}} answers on Stack Overflow (SO) at present. Users can comment on the answers to add better insights, give suggestions or reviews to improve the answer, which currently account to 76M\footnote{\url{https://data.stackexchange.com/}} comments on the platform. Several studies have been performed on the data available on Stack Overflow, such as analysing correctness of API usage in code snippets on Stack Overflow and suggesting appropriate tags for questions being posted, with an aim to improve usability of Stack Overflow \cite{zhang2018code, wang2018entagrec++, liu2018fasttagrec}.

Number of questions on SO are increasing at a rapid rate, with about 6.7K questions being posted per day\footnote{\url{shorturl.at/cKO69}}. About 2.3M\footnote{\url{https://stackoverflow.com/unanswered/tagged/?tab=noanswers}} of the total questions on Stack Overflow are unanswered. Researchers have identified that social factors such as gender, age and status of users in the community, linguistics, sentiments of the posts, and so on influence the response time for a question \cite{bosu2013building, morrison2013programming, may2019gender, silveira2019confidence, novielli2014towards}. Analysing sentiments of posts on SO could provide insights on the emotions of users, and could also motivate users to contribute to the posts \cite{novielli2014towards}. Several studies have been performed to analyse sentiment of content on social networking sites such as Twitter and Facebook. The outcomes of these studies have helped in understanding characteristics and mental health of various users on social networking sites and also helped in rumor analysis \cite{jianqiang2018deep}. Similar techniques could be used in improving user experience on Stack Overflow.



 Users of Stack Overflow who satisfy a specific criterion, currently, users with reputation 30 can post comments on answers that are not locked\footnote{\url{https://meta.stackexchange.com/questions/19756/how-do-comments-work}}. These comments specify changes and suggestions to improve the answers posted.  Most of the questions on Stack Overflow have more than one answer. Comments provide assurance of whether the given solution works or not, facilitating users to choose the best suited answer among the available ones. Augmenting these comments with emojis based on the context could motivate users to read and address these comments. Researchers have proposed techniques to enhance developer experience on Stack Overflow that include augmenting code examples with API usage examples \cite{zhang2018code}, detecting API misuse in code examples and presenting right usage patterns \cite{reinhardt2018augmenting}, augmenting questions with tags based on their context \cite{venigalla2019sotagger} and so on. However, to the best of our knowledge, we are not aware of existing work that deals with augmenting Stack Overflow with emojis based on the sentiment of posts. Hence, we propose \textit{StackEmo} as a prototype plugin to Stack Overflow, to augment comments on SO with emojis.

\section{Related Work}
\label{related}
Question and Answer websites such as Stack Overflow are being widely used by programmer hobbyists and developers as a support during programming \cite{allamanis2013and}. Researchers have proposed various ideas such as automatic tag suggestions and question templates for users who pose questions \cite{wang2018entagrec++}, API misuse warnings \cite{reinhardt2018augmenting} and API definitions \cite{venigalla2019stackdoc} to support readers of Stack Overflow. 

Beyer et al. have proposed a tool to suggest synonymous tags for a given input tag, apart of from the pre-existing synonym-tag pairs provided by Stack Overflow \cite{beyer2015synonym}. \textit{EnTagRec++} has been proposed to suggest tags to be added to the questions being posed using tag history of similar posts and user information \cite{wang2018entagrec++}. \textit{FastTagRec} uses neural networks to classify posts and suggests tags to new posts on Stack Overflow based on the classification \cite{liu2018fasttagrec}. Beyer et al. have proposed to categorized questions into seven categories based on the context in which the questions are being asked to provide insights on topics being frequently discussed \cite{beyer2019kind}. \textit{SOTagger} has been introduced to contextually tag questions on Stack Overflow into three of the six categories proposed by authors \cite{venigalla2019sotagger}. Allamanis et al. have categorized questions on Stack Overflow into 5 categories using LDA model and unsupervised ML algorithms \cite{allamanis2013and}.

Squire et al. have explored use of various question components on Stack Overflow and suggested the components to be included and excluded to post quality postings \cite{squire2014bit}. Wang et al. have presented a study aimed at understanding the factors that influence and motivate faster answering of questions being posted using a logistic regression model trained on 46 different factors. They also emphasized on the need to improve Q\&A systems to motivate answerers \cite{wang2018understanding}. \textit{SOLinker} has been proposed to identify relations among tags of questions and URLs and appropriately tag questions based on the URLs used in Q\&As using semantic analysis \cite{li2018correlation}.

\textit{Example Overflow} extracts executable code examples tangled in the text based on the user search requirements \cite{zagalsky2012example}. \textit{ExampleCheck} has been proposed as a plugin to identify API calls that are wrongly used in the code snippets on Stack Overflow. It also presents the correct usage patterns of misused APIs, thus facilitating users to understand the correct usage through examples \cite{reinhardt2018augmenting}.
\textit{StackDoc} has been proposed to augment Stack Overflow with descriptions and examples of java APIs used in questions \cite{venigalla2019stackdoc}. Zhang et al. have identified APIs being misused on Stack Overflow using \textit{ExampleCheck} \cite{zhang2018code}. Later, Zhang et al. have developed \textit{ExampleStack} to provide a list of similar code examples enumerated from Github and other Stack Overflow questions when users visit a code example on Stack Overflow \cite{zhang2019analyzing}. They observed this representation to improve confidence on an example on Stack Overflow and to help users gain awareness on the various use and misuse patterns on examples being discussed \cite{zhang2019analyzing}.

Ragkhitwetsagul et al. have identified presence of different types of toxic codes such as codes using outdated APIs and license violations, and presented an approach to identify the same \cite{ragkhitwetsagul2019toxic}.
Rahman et al. have alerted users of SO of insecure coding practices in code snippets present on Stack Overflow \cite{rahman2019snakes}. Zhang et al. have proposed \textit{DupPredictor}, as an approach to identify duplicate questions on Stack Overflow. \textit{DupPredictor} takes a new question and its parameters as input and outputs a similarity score between existing questions and the new question. This helps users who post questions to avoid re-posting existing questions \cite{zhang2015multi}. 

With the support of increased data availability aimed at providing various insights in a domain, the idea of sentiment analysis to understand various contexts and user characteristics in various platforms is being emphasized \cite{agarwal2011sentiment, guzman2014sentiment, werder2018meme}. Analysing sentiments on social networking platforms such as twitter is prevalent in the present day. Agarwal et al. have examined the sentiments of data on twitter and introduced polarity features specific to parts of speech of the tweets using a pleasantness score ranging from 1 to 3 \cite{agarwal2011sentiment}. Mohammad et al. have annotated tweets with their sentiments and emotions, using NLTK sentiment analysis tools, to observe frequency of negative and positive emotions being conveyed, and have proposed tools to automatically predict emotion in the tweets \cite{mohammad2015sentiment}. Severyn et al. have proposed a deep convolution neural network model to analyse polarity of sentiments at message and phrase level of twitter data \cite{severyn2015twitter}. Sentiments have been analysed for various components in software engineering domain such as issue reports, pull requests, commit messages \cite{werder2018meme, guzman2014sentiment}. Guzman et al. have analysed commit comments of repositories on github and provided their insights on positive and negative emotions in the commits with respect to factors such as programming languages used and number of collaborators to the repository \cite{guzman2014sentiment}. \textit{MEME} has been developed as a method to extract emotions from commits, pull requests and issues of repositories on GitHub \cite{werder2018meme}. Researchers have also emphasized the need for using better tools, trained on datasets related to software engineering to obtain better promising results \cite{lin2018sentiment, jongeling2015choosing}. 

As emphasized by Novielli et al., the idea of analysing emotions of posts on Stack Overflow is gaining importance \cite{novielli2018gold}. However, to the best of our knowledge, we are not aware of existing work that augments Stack Overflow based on emotions of the posts. Existing tools such \textit{StackDoc} \cite{venigalla2019stackdoc}, \textit{ExampleCheck} \cite{reinhardt2018augmenting} and \textit{SOTagger} \cite{venigalla2019sotagger} proposed to augment Stack Overflow, are limited to only providing further technical or contextual information based on the questions asked and solutions mentioned. Hence, we present \textit{StackEmo}, as a prototype Google Chrome plugin to augment Stack Overflow with emotions of the posts. For the purpose of prototype and considering the value of comments, we have developed \textit{StackEmo} to augment emotions of comments being displayed for each answer of questions selected by the user.
\begin{figure*}
    \centering
    \includegraphics[width = \linewidth]{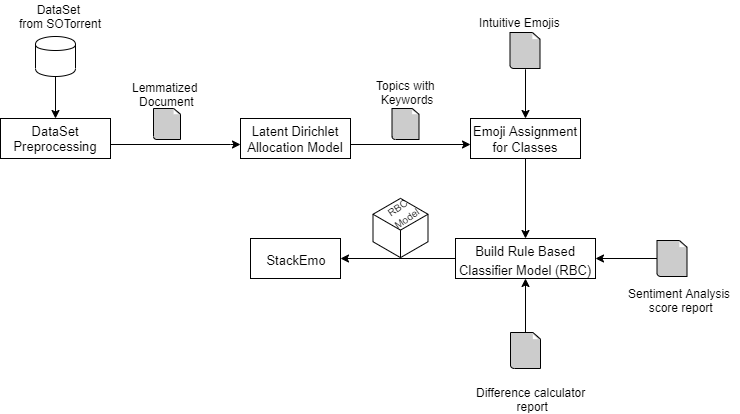}
    \caption{Design Methodology of \textit{StackEmo}}
    \label{fig:design}
\end{figure*}

\begin{figure}[!b]
    \centering
    \includegraphics[width=\linewidth, height = 4cm]{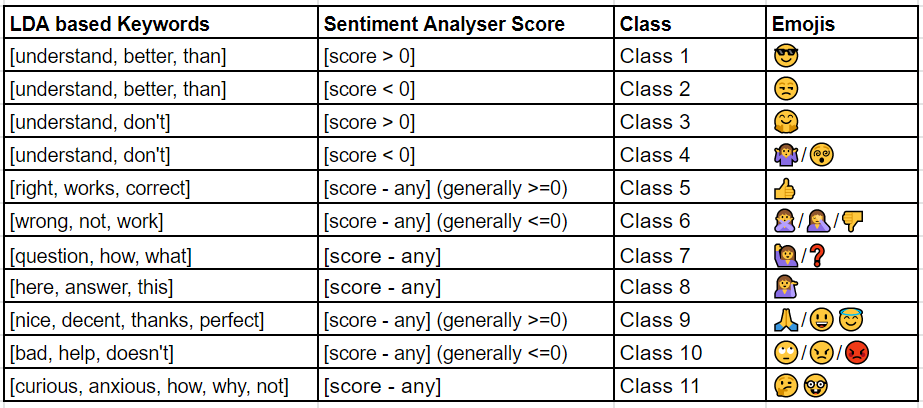}
    \caption{Keywords and Emojis of Classes}
    \label{fig:keywords}
\end{figure}

\section{Design Methodology of \textit{StackEmo}}
\label{Design}
\textit{StackEmo} aims to motivate users of Stack Overflow to read and contribute to comments given for answers on the platform by augmenting existing comments with emotions. It also aims to reduce the effort of reading through all available comments when users intend to read comments with specific emotional insights. \textit{StackEmo} appends comments on SO with emojis by characterizing the comments based on LDA clustering and Sentiment Analysis. 

While designing \textit{StackEmo}, we considered the commonly used \textit{Sentiment Analyser} library provided by NLTK. This library provides a score for a given text sentence and is commonly used by researchers sentiment analysis in twitter \cite{severyn2015twitter, mohammad2015sentiment}. This score determines positive, negative or neutral sentiment of the sentence. If the score is less than zero, it implies that the statement is negative. If the score is equal to zero, it implies that the statement is neutral. If the score is greater than zero, it implies that the statement is positive. Higher score values contribute to more positive sentiments and lower score values contribute to more negative sentiments. However, we have identified through manual inspection of comments, that scores depicting only positivity or negativity do not suffice to represent emotions. Hence, we have tried to classify comments into eleven emotion based categories based on manual introspection of posts. The choice number of classes is based on manual introspection of several comments. We have extracted Stack Overflow comments dataset from SOTorrent and passed the same to Latent Dirichlet Allocation model (LDA), with number of classes parameter equated to eleven. We have also run scripts to identify synonyms of the keywords using readily available and easy to use NLTK corpus based synonym generator library, \textit{wordnet}, and added the generated to synonyms to classes containing corresponding keywords. A difference calculator function has then been defined to identify how much a given sentence would differ with each of the defined classes.  
\textit{StackEmo} has been developed as Google Chrome plugin to Stack Overflow. We have followed a five step process, as shown in Figure \ref{fig:design} in developing \textit{StackEmo}. 

\begin{figure*}[!b]
    \centering
    \includegraphics[width=\linewidth, height = 5cm]{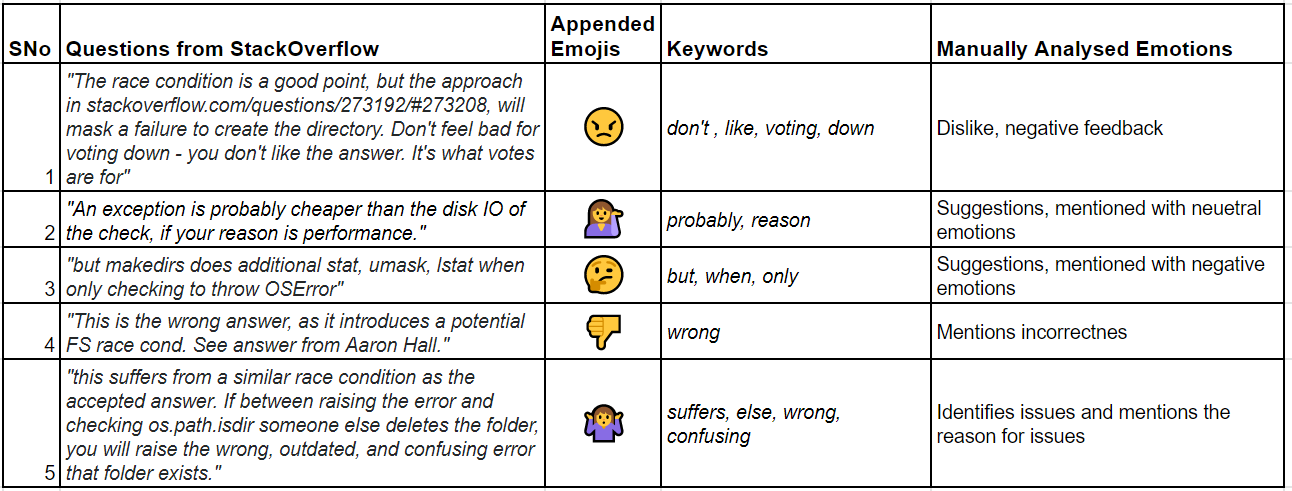}
    \caption{Example Comments extracted from Stack Overflow}
        \label{fig:example}
\end{figure*}

\textbf{Step 1 - DataSet Extraction.} We have downloaded \textit{Posts.xml} file from SOTorrent\footnote{\url{https://zenodo.org/record/2273117}} to analyse existing comments on Stack Overflow. We ramdomly selected about 50K posts from these posts as subset. We then filtered comments from the subset based on \textit{PostTypeId} column that resulted in about 30K comments. 

\textbf{Step 2 - DataSet Preprocessing.} We have preprocessed the resulting dataset of 30K comment posts for removal of stop words and punctuation and lemmatized the document using \textit{spaCy}.

\textbf{Step 3 - Latent Dirichlet Allocation Model.} We applied LDA model on the processed dataset. We specified the number of classes parameter value to be eleven. The basis for our choice of using LDA model to categorize text is the existing work of Allamanis et al. \cite{allamanis2013and} and Venigalla et al. \cite{venigalla2019sotagger} to contextually classify questions on Stack Overflow. The number of classes parameter was based on intuition and manual introspection of about 200 posts on Stack Overflow.

\textbf{Step 4 - Emoji Assignment for Classes.} We have then manually evaluated the classified posts to identify keywords in each class. Based on the resulting output categories of LDA, we prepared keyword sets for each class and labelled the classes with relevant emojis as shown in Figure \ref{fig:keywords}. 

\textbf{Step 5 - Build Rule Based Classifier Model.}We then defined a Rule Based Classifier (RBC) that takes into consideration results generated by difference calculator and NLTK sentiment predictor to classify the text passed based on predefined classification rules.

\section{Working of \textit{StackEmo}}
\label{Development}
A list of few example statements\footnote{These statements are comments extracted from \url{https://stackoverflow.com/questions/273192/how-can-i-safely-create-a-nested-directory}} on Stack Overflow, corresponding emojis that would be appended and results of manual analysis of sentiments are shown in Figure \ref{fig:example}. \textit{StackEmo} follows a six step process in appending emojis to comments on Stack Overflow, as mentioned below. 

\begin{itemize}
\item \textit{\textbf{Text Extraction-}}
When user clicks on a question on Stack Overflow, \textit{StackEmo} extracts comments of the answers to questions, if they exist.
\item \textit{\textbf{PreProcessing-}}
\begin{itemize}
    \item \textbf{Filter data-}
    The extracted text of comments is preprocessed. Any special characters present in the text are replaced by \textit{space} characters to ensure that the data extracted purely consists of alphanumeric values.
    \item \textbf{Stemming-}
    The filtered text is passed to nltk stemmer that converts each word of the text to its base form.
\end{itemize}
\item \textit{\textbf{Sentiment analysis-}}
Preprocessed text is then passed to sentiment analyser to obtain a sentiment score for the text. Score given by the sentiment analyser provides insights on how positive or negative a statement is. This resultant score is stored for further analysis.

\item \textit{\textbf{Difference Calculator-}}
The preprocessed text is passed to the difference calculator function, that compares text with words that are defined for each class. A set of keywords identified through LDA and their corresponding synonyms obtained through nltk wordnet are considered as representatives for each class. Differences of preprocessed text with each class are compared and the class with least difference score is identified.

\item \textit{\textbf{Classification-}}
The ID class identified in the previous step and the sentiment scores obtained as a result of sentiment analysis are analysed with respect to rules defined in the classifier. The posted text is thus classified into one of the eleven classes.

\item \textit{\textbf{Append Emojis-}}
Based on the output of classification, the comment post on Stack Overflow is appended with the corresponding emojis, as mentioned in Figure \ref{fig:keywords}.
\end{itemize}

\section{Evaluation}
\label{Evaluation}
\subsection{User Scenario}
\label{userscenario}
\begin{figure*}
    \centering
    \includegraphics[width=\linewidth]{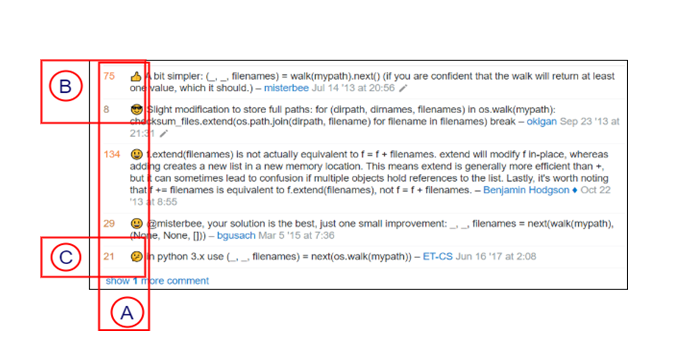}
    \caption{Snapshot of results of \textit{StackEmo} Plugin}
    \label{fig:snapshot}
\end{figure*}
\begin{figure}
    \centering
    \includegraphics[width= \linewidth]{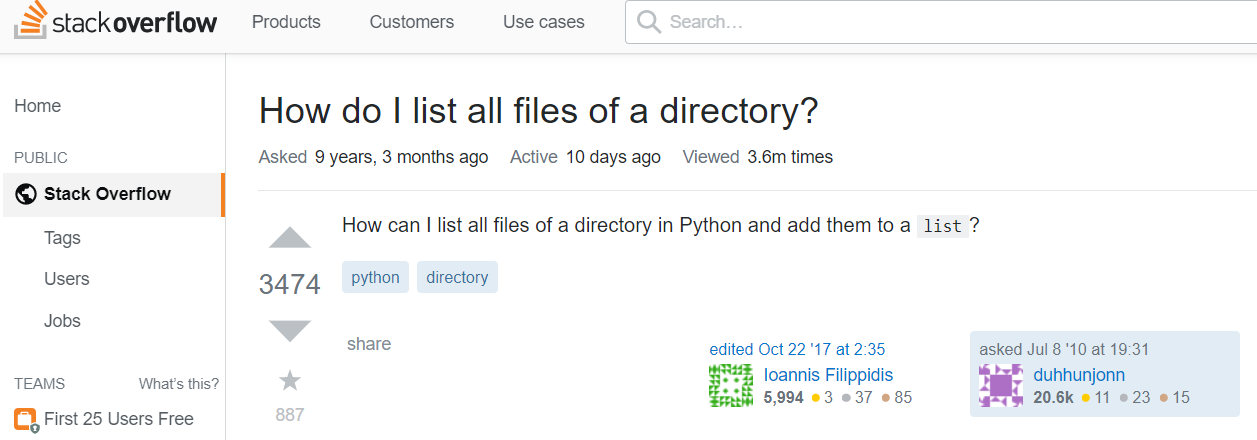}
    \caption{Question on Stack Overflow}
    \label{fig:qus}
\end{figure}
\begin{figure}
    \centering
    \includegraphics[width=\linewidth]{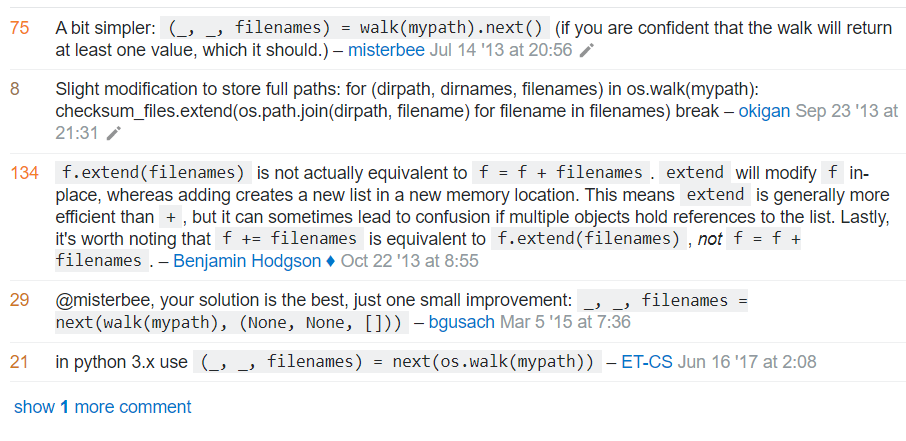}
    \caption{Comments on Stack Overflow without \textit{StackEmo}}
    \label{fig:noemo}
\end{figure}
Consider Veda to be a programmer who visits Stack Overflow with a query aimed to \textit{list all files of a directory using python}. She is then directed to the question shown in Figure \ref{fig:qus}. She goes through the first answer presented and wishes to learn more about slicing and also the authenticity of the presented answer. She scrolls down to view the comments as shown in Figure \ref{fig:noemo}. She finds comments of the posts, but finds it time consuming to read through all the comments present. She then adds \textit{StackEmo} as an extension to Google Chrome web browser and reloads the question page. She is then displayed with comments that are appended with emojis, as shown in [A] of Figure \ref{fig:snapshot}. Veda then reads through the comments that are intended to enhance the answer, based on the emojis that belong to Class 1 and Class 5, that imply positive suggestions, specifically referring to a solution that \textit{works} and that the solution \textit{helps in improving the user understanding}, as shown in [B] of Figure \ref{fig:snapshot}. She conveniently ignores comments augmented with emojis assigned for Class 11 (shown in [C] of Figure \ref{fig:snapshot}), as it is perceived to deal with further questions, displaying curiosity and anxiousness.

\subsection{User Survey}
\label{Survey}
We followed similar approach of in-user survey used by Zhang et al. in \cite{zhang2019analyzing} to evaluate \textit{StackEmo} with 30 university students in the age group 18 to 23 years. 28 of them are undergraduate computer science students and the remaining two participants are post graduate students of computer science, currently in their third semester (age group of 21-23 years). 25 among the 28 participants are in their fifth semester (age group of 19-20 years), 2 participants in their seventh semester (age group of 20-21 years) and one participant is in her third semester (age group of 18-19 years) of under graduation currently. 

All the volunteers were requested to install \textit{StackEmo} and analyse the emoji-augmented comments. A clear documentation that specifies location which \textit{StackEmo} could be downloaded, steps to be followed to install it as chrome plugin, idea being demonstrated by the plugin and steps to be followed to evaluate the plugin, is sent to volunteers. An in-lab survey has been conducted for 25 undergraduate participants and 2 post graduate participants for a duration of 15 minutes. An email was communicated to rest of the three participants. They were instructed to follow the steps mentioned in the shared document to assess \textit{StackEmo}. As a part of the survey, all the participants were instructed to freely browse comments on SO and review the appended emojis. Then, the participants were requested to navigate to user survey link provided and answer the questions using 5-point likert scale. Questionnaire circulated among the volunteers is shown in Table \ref{tab:table2}. 
\begin{table}
    \caption{Questions in survey using a 5-point Likert Scale.}
    \label{tab:table2}
    \centering
    \begin{tabular}{|l|}
     \hline
     \textbf{Q1:} How did you like the plug-in interface?\\(5=strongly like, 1=strongly dislike)\\
   \\
   
     \textbf{Q2:} According to you, \textit{StackEmo} has rendered emotion\\ of posts satisfactorily, with appropriate emojis.\\ (5=strongly agree, 1=strongly disagree)\\
   \\
     \textbf{Q3:} \textit{StackEmo} has helped me in identifying useful\\ comments through emojis.\\
     (5=strongly agree, 1=strongly disagree)\\
    \\
  
      \textbf{Q4:} \textit{StackEmo} has motivated me to view comments,\\ which helped in getting better insights\\ about an answer. \\(5=strongly agree, 1=strongly disagree)\\
    \\
    \textbf{Q5:} I will recommend \textit{StackEmo} to my peers.\\ (5=strongly agree, 1=strongly disagree)\\
     \hline
 \end{tabular}
\end{table}

\section{Results}
\label{Results}
\begin{figure}
    \centering
    \includegraphics[width = \linewidth]{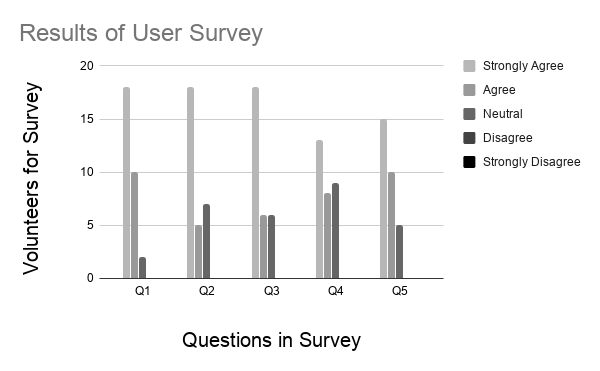}
    \caption{Results of \textit{StackEmo} user survey}
    \label{fig:res}
\end{figure}

To evaluate the survey responses, we have given weighted values for each point of the Likert Scale, as shown in Table \ref{tab:table2}. The average score for each question is then considered as the evaluation metric of \textit{StackEmo} based on user responses. 
As reported in Fig \ref{fig:res}, 85\% of the participants liked the interface of \textit{StackEmo}. In Q2, 76.8\% of participants have agreed that \textit{StackEmo} has satisfactorily rendered emotions to posts, with appropriate emojis. However, 3 of the participants reported instances of emojis not being appropriate for the comment. The scores in Q3 and Q4 infer that \textit{StackEmo} has helped volunteers to identify useful comments and motivated them to view comments and get better insights (scores: 80\% in Q3 and 79.3\% in Q4). Participants have mentioned their suggestion to enhance \textit{StackEmo} by reconfiguring the emojis. In Q5, 25 of 30 participants have agreed to recommend \textit{StackEmo} to their peers (score:83\%). Two of the participants also pointed out that \textit{StackEmo} fails to render emojis when users clicked on \textit{"show more comments"} link.

\section{Threats to Validity}
\label{threats}
The prototype version of \textit{StackEmo} has been developed as a first step towards augmenting Stack Overflow posts with emotions being conveyed by the posts.
\textit{StackEmo} classifies the comments into one of the eleven proposed classes and augments the comments on Stack Overflow with predetermined emojis.

The keywords specified for each class are based on manual analysis of top 20 comments for each class, generated by LDA model. The rank of these comments depend on probability values generated by the model for comments belonging to each class. Classification based on the number of comments inspected, might result in inaccuracies, considering the huge amount of data available. To validate the correctness of classification, we have manually analysed emotions of 30 random comments on Stack Overflow. We observed that about 8 comments were tagged with different emojis than expected. 

LDA model was designed to classify a comment into one of the eleven categories. The choice number of categories was based on intuition and opinions of 2 developers, after a random walkthrough of posts on Stack Overflow. A more concrete approach in deciding the number of classes shall be explored in future versions.

The correctness of sentiment scores obtained is dependant on efficiency of Sentiment Analyser provided by NLTK. Researchers have also mentioned the inefficiency of nltk sentiment analyser in the area of software engineering. There is thus, a need to adopt better analysers to analyse sentiment of posts present on Stack Overflow and other coding platforms. Also. the keyword synonyms are generated by synonym generator of wordnet library provided by NLTK. As a result, few synonyms might not have been considered, if not reported by the generator, and few might be out of context in view of the domain being considered. The choice of emojis is also based on general knowledge and acceptance of the purpose of emojis, which we plan to revise in the next version.

\section{Conclusion and Future Work}
Considering the increasing importance of sentiment analysis in various domains of software engineering and the effort being involved by programmers in learning through comments on Stack Overflow, we proposed an approach to augment Stack Overflow comments with their corresponding emotions. In this paper, we presented prototype version of \textit{StackEmo}, a Google Chrome plugin. \textit{StackEmo} augments comments on Stack Overflow by appending them with emojis based on the sentiment of comments. The comments presented on Stack Overflow, for an answer are extracted and processed, when a question is being viewed on Stack Overflow. The processed comments are then analysed for the emotion being conveyed and classified into one of the predefined eleven classes. An emoji is then appended to the comment based on predefined rules and displayed on Stack Overflow. \textit{StackEmo} thus aims to provide insights about the comments, and could help users in reading through comments that only express a specific emotion, or avoiding comments that express specific emotions. Presence of emojis on the platform might motivate the users to learn better by reviewing more comments and also might motivate practitioners to address more number of comments. However, the plugin has to be further evaluated to be able to comment on motivation factors of users. We have evaluated \textit{StackEmo} with 30 university students, using a 5 point likert scale based questionnaire, through a in-user survey. 25 of the participants reported positive experience with the plugin and agreed to recommend \textit{StackEmo} to their peers. Participants have also suggested to add emojis for questions and answers on Stack Overflow. 

As a part of future work, we plan to improve the sentiment analysis by considering analysers specific to software engineering domain. We also plan to conduct a large scale user study to arrive at a consensus on the number of classes and features of these classes. We shall improve the accuracy of \textit{StackEmo}, by performing extensive manual analysis on posts classified using LDA model, to identify keywords for each of the classes. We plan to investigate more concrete approaches to classify the posts based on emotions being conveyed. As suggested by the participants of user survey, we also plan to extend \textit{StackEmo} to augment question and answer posts with emojis based on the sentiment of the posts.

\begin{acks}
We thank volunteers for their valuable time and honest feedback in evaluating \textit{StackEmo}. We thank the undergraduate Vagavolu Dheeraj for helping us in developing the plugin. 
\end{acks}

\bibliographystyle{ACM-Reference-Format}
\bibliography{references.bib}

\end{document}